\documentstyle[12pt]{article}
\def\singlespace {\smallskipamount=3.75pt plus1pt minus1pt
                  \medskipamount=7.5pt plus2pt minus2pt
                  \bigskipamount=15pt plus4pt minus4pt
                  \normalbaselineskip=15pt plus0pt minus0pt
                  \normallineskip=1pt
                  \normallineskiplimit=0pt
                  \jot=3.75pt
                  {\def\smallskip {\vskip\smallskipamount}}
                  {\def\medskip   {\vskip\medskipamount}}
                  {\def\bigskip   {\vskip\bigskipamount}}
                  {\setbox\strutbox=\hbox{\vrule 
                    height10.5pt depth4.5pt width 0pt}}
                  \parskip 7.5pt
                  \normalbaselines}

\begin{document}

\textheight 9.0in \topmargin -0.5in \textwidth 6.5in \oddsidemargin -0.01in

\bigskip

\begin{center}
{\bf {\Large Quantum Mechanics without Spacetime II}}\\
{- {\it Noncommutative geometry and the free point particle} -}
\end{center}

\medskip
\centerline{\bf T. P. Singh}
\smallskip
\centerline{\small {\it Tata Institute of Fundamental Research}}
\centerline{\small {\it Homi 
Bhabha Road, Mumbai 400 005, India}} 
\centerline{{\small \tt email address: tpsingh@tifr.res.in}}

\bigskip

\centerline{\bf Abstract}
\noindent{\small
In a recent paper we have suggested that a formulation of quantum mechanics should exist, which does not require the concept of time, and that the appropriate mathematical language for such a formulation is noncommutative differential geometry. In the present paper we discuss this formulation for the free point particle, by introducing a commutation relation for a set of noncommuting coordinates. The sought for background independent quantum mechanics is derived from this commutation relation for the coordinates. We propose that the basic equations are invariant under automorphisms which map one set of coordinates to another- this is a natural generalization of diffeomorphism invariance when one makes a transition to noncommutative geometry. The background independent description becomes equivalent to standard quantum mechanics if a spacetime manifold exists, because of the proposed automorphism invariance. The suggested basic equations also give a quantum gravitational description of the free particle.}

\bigskip

\singlespace

\section{Introduction}

In a recent paper \cite{tps} we have argued that there should exist a formulation of quantum mechanics which does not require the concept of time. This has been suggested because, as is well-known, spacetime manifold and its 
pseudo-Riemannean geometry are classical concepts valid only when classical matter sources are present to produce them. In \cite{tps} we called such a timeless formulation Fundamental Quantum Mechanics, although a better name perhaps is Background Independent Quantum Mechanics (BIQM).

In order to construct a BIQM, the simplest case to consider is the free point particle of mass $m$. As discussed in \cite{tps}, a background independent description of the quantum free particle must include also a description of the gravitational field of the particle. Such a description is hence naturally a quantum gravity theory for the free particle. In the limit that the particle's mass $m$ is much larger than Planck mass $m_{p}$, the theory reduces to classical general relativity on a spacetime manifold, and the classical equation of motion for the particle. On the other hand, in the limit $m\ll m_{p}$ the gravitational effects of the particle can be neglected and the theory reduces to the above-mentioned BIQM.

We have proposed in \cite{tps} that the appropriate language for describing the free particle in quantum gravity is noncommutative differential geometry. In the present paper we sketch the basic equations for the free particle, by introducing a set of structure functions for the noncommuting coordinates of the particle. It is also suggested that the BIQM, as well as the gravitational effects of the particle, are to be derived from these structure functions. We also propose that the symmetry of this theory is invariance under automorphisms which map one set of coordinates to another. In particular, BIQM can be transformed to standard quantum mechanics if there exists classical matter in the Universe which gives rise to a classical spacetime manifold.

Our construction is highly non-rigorous, in so far as a precise application of noncommutative geometry to the physical problem at hand is concerned. However, the physical picture presented here appears to be on the right track, and we hope to make the mathematical development more concrete in a forthcoming investigation.

The idea that there should exist a time-independent formulation of quantum mechanics is not new. Previous studies include the work of Hartle \cite{Har} on the sum-over-histories generalization of quantum mechanics, the work of Rovelli and collaborators on developing a quantum mechanics without time \cite{Rov},
\cite{Mon}, and the work of Kanatchikov \cite{Kan}. The discussion given in the present series of papers is probably the first attempt to use noncommutative geometry for constructing a spacetimeless quantum mechanics. A recent paper by Corichi et al. \cite{Cor} is quite similar, in spirit, to the motivation and the approach of the present paper towards a quantum theory of gravity. In the context of string theory, a dynamical generation of spacetime in the weak coupling limit of a matrix model has been proposed in \cite{ikkt}.

\section{The case of the free point particle}

Consider the case that there is only one particle in the Universe, and that its mass is so small compared to Planck mass that it is not possible to talk of a background spacetime manifold. In order to describe the quantum mechanics and gravity of this point particle,
we introduce a set of noncommuting coordinates $x^{i}$ satisfying the commutation relations
\begin{equation}
                    [x^{i}, x^{j}]=iA^{ij}_{k}x^{k}     \label{comm}
\end{equation}
where the $A^{ij}_{k}$ are constant structure functions. From these relations one deduces a noncommutative differential calculus
\begin{equation}
                    [x^{i}, dx^{j}]=iB^{ij}_{k}dx^{k} \label{qm}
\end{equation}
where
\begin{equation}
                   A^{ij}_{k}=B^{ij}_{k}-B^{ji}_{k}.
\label{abb}
\end{equation}
                    
Next, we introduce a connection and a corresponding covariant derivative 
$\nabla^{i}$ satisfying the relation
\begin{equation}
                      [\nabla^{i},\nabla^{j}]V^{k}=C^{ijk}_{l}V^{l}.
\end{equation}

Equations (1-4), along with an equation of `motion' (to be discussed in the next Section)
describe the Background Independent Quantum Gravity for the free particle. The unknown
functions $A^{ij}_{k}$ and $C^{ijk}_{l}$ are to be determined by the mass of the particle, 
in the spirit of general relativity. In the limit $m\gg m_{p}$ these equations should reduce
to classical general relativity on the spacetime manifold with coordinates $x^{i}$, where the noncommutativity of the 
coordinates becomes negligible. In the limit $m\ll m_{p}$ they reduce to the 
Background Independent Quantum Mechanics (BIQM), where the gravitational field of
the particle can be ignored. It then has to be shown that BIQM is
equivalent to standard quantum mechanics whenever a background spacetime is available.

We propose to determine the connection by suggesting a physical relation between the functions $C^{ijk}_{l}$ and the structure functions $A^{ij}_k$ in
Eqn.(\ref{comm}), as follows. In the limit that the mass of the particle is much larger than Planck mass, $C^{ijk}_{l}$ should become the Riemann curvature tensor, $R^{ijk}_{l}$, which according to Einstein equations has a typical component
\begin{equation}
R^{ijk}_{l} \sim {Gm \over R^{3}c^{2} }\sim {R_{S}\over R^{3}} \sim  
{L_{p}\over R^{3}} 
\left( m\over m_{p} \right) .           \label{sch}
\end{equation}  

The structure functions $A^{ij}_{k}$ have the dimension of length and we propose that in this large mass limit a typical component $A$ of $A^{ij}_{k}$ is related to a typical component $C$ of $C^{ijk}_{l}$ as
\begin{equation} 
                 A \sim {L_{p}^{2}\over R^{3}} {1\over C} \sim 
                 L_{p} \left( m_{p}\over m \right) \sim {\hbar\over mc}.
\label{ac}
\end{equation}
Our proposal to relate the $A$'s to the $C$'s is at this stage ad hoc but it is interesting nonetheless that a typical component for $A$ can be made to come out as the Compton wavelength of the particle. Such a relation between the $A$'s and the $C$'s could also be expected on the basis of the striking duality in the definitions of the Schwarzschild radius $R_{S}$ and the Compton wavelength $\lambda_c$ of the particle: $R_{S}\sim L_{p}(m/m_{p})$ and $\lambda_c \sim L_{p} (m_{p}/m)$. 

If we take (\ref{abb}) to mean that a typical component $B$ of $B^{ij}_{k}$ is also of the order $\hbar/mc$ then (\ref{qm}) may be written as
\begin{equation}
                    [x^{i}, m\ dx^{j}] \sim i{\hbar\over c}
                      \ I^{ij}_{k}dx^{k}  \label{fqm}
\end{equation}
where the $I^{ij}_{k}$ are dimensionless quantities of order unity. This equation should be regarded as the spacetimeless equivalent of the standard commutation relation $[q,p]=i\hbar$. As a naive example, a `division' by $dx^{0}/c= dt$ might allow the left hand side to be thought of as a generalisation of $[q,p]$ in the noncommutative context, and the right hand side as a correction to $i\hbar$. In particular, it is possible that for $i=j$ we get exactly
\begin{equation}
                    [x,p]=i\hbar              \label{ncr}
\end{equation}
in the spacetimeless picture provided by the noncommuting coordinates $x^{i}$.

Since this choice for the $A$'s gives the expected structure for the free 
particle's quantum mechanics, we assume that such a choice holds also in the small
mass limit $m\ll m_{p}$. The chosen $A$'s then determine the connection and the functions
$C^{ijk}_{l}$ via a relation similar to Eqn. (6). Of course, it is not meaningful
in the small mass case to talk of a physical distance $R$; instead it may be reasonable that
$R$ gets replaced by $L_{p}$ in this limit.

A more systematic way of determining the $A$'s would be to write the r.h.s. of (\ref{comm}) as
\begin{equation}
\theta^{\mu\nu}=A^{\mu\nu}_{\eta}x^{\eta}.
\end{equation}
$\theta^{\mu\nu}$ should then be determined by $m$, via a new field equation, such that in the large mass limit $m\gg m_{p}$, a typical component $\theta$ grows linearly with distance as $R$. This could then imply the constancy of $A$. 
In terms of  $\theta$, Eqn.(\ref{ac}) can be written more suggestively, as
\begin{equation}
   {R\over L_{p}}\theta \sim \left({R\over L_{p}}C\right)^{-1}.
\end{equation}

\section{Relation with standard quantum mechanics}

We next address the important question of relating this `Background Independent Quantum Mechanics' to the standard quantum mechanics.
As noted above, standard quantum mechanics should emerge when classical matter is also
present to endow the Universe with a classical spacetime manifold structure. Consider,
for the sake of simplifying the discussion, that there is only one other particle, having a
mass $m_{2}\gg m_{p}$. The basic coordinates $x_{2}^{i}$ associated with this particle are
very nearly commuting, and they provide the classical spacetime manifold with respect to
which the standard quantum mechanics of our basic particle $m$ is written.

The transformation from the noncommuting coordinates of $m$ to the commuting coordinates of $m_{2}$ is
via an automorphism, which is a natural generalization of diffeomorphisms when one
makes a transition to noncommutative differential geometry. We propose that the basic equations
(1-4) introduced above are invariant under automorphisms - they retain their form even
when we use the spacetime coordinates of $m_{2}$ to describe the motion of particle $m$.
It is in this sense that the BIQM is related to standard quantum mechanics. We can
describe quantum mechanics of $m$ either by using the noncommuting coordinates of $m$,
or the very nearly commuting coordinates of $m_{2}$. The former approach gives BIQM, and
the latter gives the more familiar description on a spacetime background. Of course BIQM
is more fundamental, not having to depend on an external classical system. Furthermore, the
description in terms of noncommuting coordinates also incorporates the quantized 
gravitational field of $m$.

One could give another reason for introducing invariance under automorphisms. If one considers enlarging the symmetry group of general relativity (i.e. general coordinate transformations or diffeomorphisms), the next step could be to demand that these coordinates themselves be quantum-mechanical and noncommuting. The invariance under diffeomorphisms is then replaced by invariance under automorphisms. However, the rule of noncommutation for the coordinates cannot be borrowed from or determined by standard quantum mechanics, as the latter already assumes a classical spacetime background. Instead, the noncommutation rule for coordinates has to be written {\it ab\ initio}, as for instance proposed above in Eqn. (1), and standard quantum mechanics is to be derived as a consequence. The requirement that the theory be invariant under automorphisms also appears to open up a powerful avenue for unification of gravity with other interactions, as discussed by Connes (see next Section).

An important difference between our proposal and earlier studies of noncommutative spacetimes is that previous studies suggest that spacetime becomes noncommutative at a sufficiently small length like Planck scale. What we are proposing, motivated by our search for a spacetimeless description of quantum mechanics, is that there is a set of noncommuting coordinates associated with every quantum mechanical particle, and this is a priori independent of Planck length scale physics.

\section{Discussion}

In standard quantum mechanics, the physical state of the system can be thought of as the noncommutative analog of a derivation (vector field) in phase space. Taking this as a clue we suggest that in the BIQM proposed in this paper, the state of the free particle is a derivation in the space of the noncommuting coordinates. In the large mass commutative limit, this state becomes a classical spacetime trajectory (since one of the coordinates becomes time).

Next, one has to address the question of the equation of `motion' satisfied by the physical state. One could not really call it an equation of motion, since there is no evolution in the usual sense. Rather, it is more like a field equation satisfied by the physical state, which becomes equivalent to the standard quantum mechanical equation of motion when one transforms, via an automorphism, to the spacetime coordinates made available by classical matter.

Our proposal for such a field equation is strongly motivated by the definition of distance given by Connes \cite{cond} for a noncommutative geometry. According to Connes, the infinitesimal distance $ds$ between two neighbouring states is the inverse of the Dirac operator $D$ (i.e. it is the fermion propagator $D^{-1}$):
\begin{equation}
ds=D^{-1}           \label{dis}
\end{equation}

This suggests that the equation of `motion' for the free particle in BIQM should be the Dirac equation in the noncommuting coordinates given by (\ref{comm}). This is also suggested by the requirement of invariance under automorphisms: if on a standard spacetime background the equation of motion is the Dirac equation, then the equation of motion in the noncommutative case should be obtained by replacing the commuting coordinates by noncommuting ones. The spirit here is similar to the situation when one goes to a curved space equation from a flat space equation: replace ordinary derivatives by covariant derivatives.

The equation of motion in noncommuting coordinates may have new features not observed in the commuting case, including perhaps mass quantization. We hope to investigate these issues in the near future. In view of Connes' definition of distance it could also be conjectured that quantum mechanical evolution, as given say by the usual path integral propagator, represents `geodesic' motion in the underlying noncommutative geometry given by Eqn. (\ref{comm}).

We close with three general remarks. Firstly, in previously considered applications of noncommutative differential geometry to gravitational physics, one assumes that classical spacetime becomes noncommuting at the Planck scale. This leaves open the (yet unresolved) issue of quantizing this noncommutative spacetime. The picture we have presented here is very different - for us, the noncommutative spacetime is intrinsically quantum mechanical. This helps develop a purely geometric description of quantum mechanics; we do not quantize a classical system, but the system given by Eqns. (1-4) is quantum mechanical to begin with. This echoes a view that emerges from M-theory  - the known dualities between classical string theories and quantum ones suggest that our quantum theories should be quantum to start with. Also, the `UV-IR' connection between Eqn. (1) and Eqn. (4) might be of some use in understanding the holographic principle.

Secondly, the work of Connes and collaborators provides a very suggestive path to unification. As discussed by Connes \cite{cond}, the symmetry group of the Einstein lagrangian plus the standard model Lagrangian is the semi-direct product of the diffeomorphism group and the group of gauge transformations. This symmetry group cannot be the diffeomorphism group of some new space. However, if one allows the space to be noncommutative, there is a noncommutative algebra whose group of internal automorphisms corresponds to gauge transformations, and the quotient of automorphisms with respect to the internal automorphisms corresponds to diffeomorphisms. An open issue has been that of quantization. If, though, one were to invoke the `intrinsically quantum mechanical' algebra of coordinates as for instance given by Eqn. (1), there appears to be a possibility of developing a quantized unification along the lines proposed by Connes and collaborators.

Lastly, we quote an observation of Connes \cite{cond}:

``Noncommutative measure spaces evolve with time. In other words there is a `god-given' one parameter group of automorphisms of the algebra $M$ of measurable coordinates. It is given by the group homomorphism
\begin{equation}
        \delta: \Re\rightarrow {\rm Out(M)=Aut(M)/Int(M)}
\end{equation}
from the additive group $\Re$ to the group of automorphism classes of $M$ modulo inner automorphisms.'' 
 
In the context of the algebra of coordinates proposed in the present paper, we could ask if the above observation of Connes suggests the existence of a `time' in quantum gravity, from which the usual notion of time emerges in the classical limit.

{\bf ACKNOWLEDGEMENTS:} It is a pleasure to thank K. P. Yogendran for useful discussions. I acknowledge the partial support of FCT, Portugal under contract number SAPIENS/32694/99.

\end{document}